\documentclass[%
 reprint,
superscriptaddress,
 amsmath,amssymb,
 aps,
]{revtex4-1}

\usepackage{graphicx}

\usepackage{dcolumn}
\usepackage{bm}

\usepackage{multirow}
\usepackage{color}

\newcommand{\appropto}{\mathrel{\vcenter{
  \offinterlineskip\halign{\hfil$##$\cr
    \propto\cr\noalign{\kern2pt}\sim\cr\noalign{\kern-2pt}}}}}
    
\let\originalleft\left
\let\originalright\right
\renewcommand{\left}{\mathopen{}\mathclose\bgroup\originalleft}
\renewcommand{\right}{\aftergroup\egroup\originalright}

\usepackage{braket}

\usepackage{mathtools}

\DeclarePairedDelimiter\abs{\lvert}{\rvert}%
\DeclarePairedDelimiter\norm{\lVert}{\rVert}%

\makeatletter
\let\oldabs\abs
\def\abs{\@ifstar{\oldabs}{\oldabs*}}
\let\oldnorm\norm
\def\norm{\@ifstar{\oldnorm}{\oldnorm*}}
\makeatother

\newcommand{\goodchi}{\protect\raisebox{2pt}{$\chi$}}

\begin{document}

\preprint{APS/123-QED}

\title{Characterization of the proposed $4\textrm{-}\alpha$ cluster state\\ candidate in $\mathrm{^{16}O}$}

\author{K.C.W.~Li}
\affiliation{Department of Physics, University of Stellenbosch, Private Bag X1, 7602 Matieland, South Africa}
\affiliation{iThemba LABS, National Research Foundation, PO Box 722, Somerset West 7129, South Africa}

\author{R.~Neveling}%
\affiliation{iThemba LABS, National Research Foundation, PO Box 722, Somerset West 7129, South Africa}

\author{P.~Adsley}%
\affiliation{Department of Physics, University of Stellenbosch, Private Bag X1, 7602 Matieland, South Africa}
\affiliation{iThemba LABS, National Research Foundation, PO Box 722, Somerset West 7129, South Africa}

\author{P.~Papka}%
\affiliation{Department of Physics, University of Stellenbosch, Private Bag X1, 7602 Matieland, South Africa}
\affiliation{iThemba LABS, National Research Foundation, PO Box 722, Somerset West 7129, South Africa}

\author{F.D.~Smit}%
\affiliation{iThemba LABS, National Research Foundation, PO Box 722, Somerset West 7129, South Africa}

\author{J.W.~Br\"ummer}%
\affiliation{Department of Physics, University of Stellenbosch, Private Bag X1, 7602 Matieland, South Africa}

\author{C.~Aa.~Diget}%
\affiliation{Department of Physics, University of York, York, United Kingdom}

\author{M.~Freer}%
\affiliation{School of Physics and Astronomy, University of Birmingham, Edgbaston Birmingham, B15 2TT, United Kingdom}

\author{M.N.~Harakeh}%
\affiliation{University of Groningen, KVI Center for Advanced Radiation Technology, 9700 AB Groningen, The Netherlands}

\author{Tz.~Kokalova}%
\affiliation{School of Physics and Astronomy, University of Birmingham, Edgbaston Birmingham, B15 2TT, United Kingdom}

\author{F.~Nemulodi}%
\affiliation{iThemba LABS, National Research Foundation, PO Box 722, Somerset West 7129, South Africa}

\author{L.~Pellegri}%
\affiliation{iThemba LABS, National Research Foundation, PO Box 722, Somerset West 7129, South Africa}
\affiliation{University of the Witwatersrand, Johannesburg Wits 2050, South Africa}

\author{B.~Rebeiro}%
\affiliation{Department of Physics, University of the Western Cape, Bellville ZA-7535, South Africa}

\author{J.A.~Swartz$^*$}%
\affiliation{KU Leuven, Instituut voor Kern- en Stralingsfysica, Celestijnenlaan 200D, B-3001 Leuven, Belgium}
\thanks{Present institution: Department of Physics and Astronomy, University of Aarhus, DK-8000 Aarhus C, Denmark}%

\author{S.~Triambak}%
\affiliation{Department of Physics, University of the Western Cape, Bellville ZA-7535, South Africa}

\author{J.J.~van~Zyl}%
\affiliation{Department of Physics, University of Stellenbosch, Private Bag X1, 7602 Matieland, South Africa}

\author{C.~Wheldon}%
\affiliation{School of Physics and Astronomy, University of Birmingham, Edgbaston Birmingham, B15 2TT, United Kingdom}


\date{\today}

\begin{abstract}

\noindent The $\mathrm{^{16}O}(\alpha, \alpha^{\prime})$ reaction was studied at $\theta_{lab} = 0^\circ$ at an incident energy of $\textrm{E}_{lab}$ = 200 MeV using the K600 magnetic spectrometer at iThemba LABS. 
Proton and $\alpha$-decay from the natural parity states were observed in a large-acceptance silicon-strip detector array at backward angles. 
The coincident charged particle measurements were used to characterize the decay channels of the $0_{6}^{+}$ state in $\mathrm{^{16}O}$ located at $E_{x} = 15.097(5)$ MeV.
This state is identified by several theoretical cluster calculations to be a good candidate for the 4-$\alpha$ cluster state.
The results of this work suggest the presence of a previously unidentified resonance at $E_{x}\approx15$ MeV that does not exhibit a $0^{+}$ character.
This unresolved resonance may have contaminated previous observations of the $0_{6}^{+}$ state.


\end{abstract}

\maketitle




\noindent Light nuclei are expected to exhibit cluster-like properties in excited states with a low density structure.
Such states should exist particularly at excitation energies near the separation energies to these clusters, as described by the 
Ikeda diagram \cite{Ikeda01011980}. 
The Hoyle state, the $\mathrm{0_{2}^{+}}$ state at $E_{x} = 7.654$ MeV in $\mathrm{^{12}C}$, is considered the archetype of a state that 
exhibits $\alpha$-particle structure, with one option being a 3$\alpha$ gas-like 
structure similar to a Bose-Einstein condensate consisting of three $\alpha$-particles all occupying the lowest 0S state \cite{Schuck-2013}.

It is expected that equivalent Hoyle-like states should also exist in heavier $\alpha$-conjugate nuclei such as 
$\mathrm{^{16}O}$ and $\mathrm{^{20}Ne}$ \cite{PhysRevC.74.044311}. 
Indeed a potential candidate in $\mathrm{^{16}O}$ has been identified. 
Funaki {\it et al.} \cite{PhysRevLett.101.082502} solved a four-body equation of motion based on the Orthogonality Condition Model (OCM) 
that succeeded in reproducing the observed $0^{+}$ spectrum in $\mathrm{^{16}O}$ up to the $0_{6}^{+}$ state. 
It was suggested that the 4$\alpha$ condensation state could be assigned to the $0_{6}^{+}$ state at $E_{x} = 15.097(5)$ MeV 
(see Table \ref{fig:/table/PreviousResults}).
The $0_{6}^{+}$ state obtained from the calculation is 2 MeV above the four $\alpha$-particle breakup threshold 
($\textrm{S}_{4\alpha}$ = 14.437 MeV) and has a large radius of 5 fm, indicating a dilute density structure. 
Ohkubo and Hirabayashi showed in a study of $\alpha$+$\mathrm{^{12}C}$ elastic and inelastic scattering \cite{Ohkubo2010} 
that the moment of inertia of the $0_{6}^{+}$ state is drastically reduced, which suggests that it is a good candidate for 
the 4$\alpha$ cluster condensate state.
Calculations performed with the Tohsaki-Horiuchi-Schuck-R\"opke (THSR) $\alpha$-cluster wave function \cite{PhysRevC.82.024312} 
also support this notion with an estimated total width of 34 keV for the $0_{6}^{+}$ state \cite{PhysRevC.80.064326}, much smaller 
than the experimentally determined value of 166(30) keV \cite{TILLEY19931}.

\begin{table}[b]
\begin{ruledtabular}
\begin{tabular}{ l  l  l  l  l }
      	\multirow{2}{*}{Reference}	&\multirow{2}{*}{Year}&	$E_{R}$			&	Width			&	\multirow{2}{*}{Reaction}	 	\\
      						&				&	 {[}MeV{]} 			&	{[}keV{]}			&	\vspace{2pt}\\
   \hline
\rule{0pt}{9.5pt}NPA 180 282 	& 1972 	&	15.17(5)			&	190(30)			&	$\mathrm{^{12}C}(\alpha, \alpha^{\prime})$				\\
	NPA 294 161 			& 1978	&	15.10(5)			&	327(100)			&	$\mathrm{^{15}N}(p, \alpha), \mathrm{^{15}N}(p, p^{\prime})$	\\
	NPA 305 63 			& 1978	&	15.103(5)			&	-				&	$\mathrm{^{14}N}(\mathrm{^{3}He}, p)$							\\
	PRC 25 729 			& 1982	&	15.066(11)		&	166(30)			&	$\mathrm{^{12}C}(\alpha, \alpha^{\prime}), \mathrm{^{15}N}(p, \alpha)$	\\
	NNDC				& 2016	&	15.097(5)			&	166(30)			&	-														\\
	This work				& 2016	&	15.076(7)			&	162(4)			&	$\mathrm{^{16}O}(\alpha, \alpha^{\prime})$						\\
\end{tabular}
\end{ruledtabular}
   \caption[Literature values for the $0_{6}^{+}$ state in $\mathrm{^{16}O}$]{Literature values for the $0_{6}^{+}$ state in $\mathrm{^{16}O}$.}
   \label{fig:/table/PreviousResults}
\end{table}

\begin{figure*}[!t]
\centering
\includegraphics[width=\textwidth]{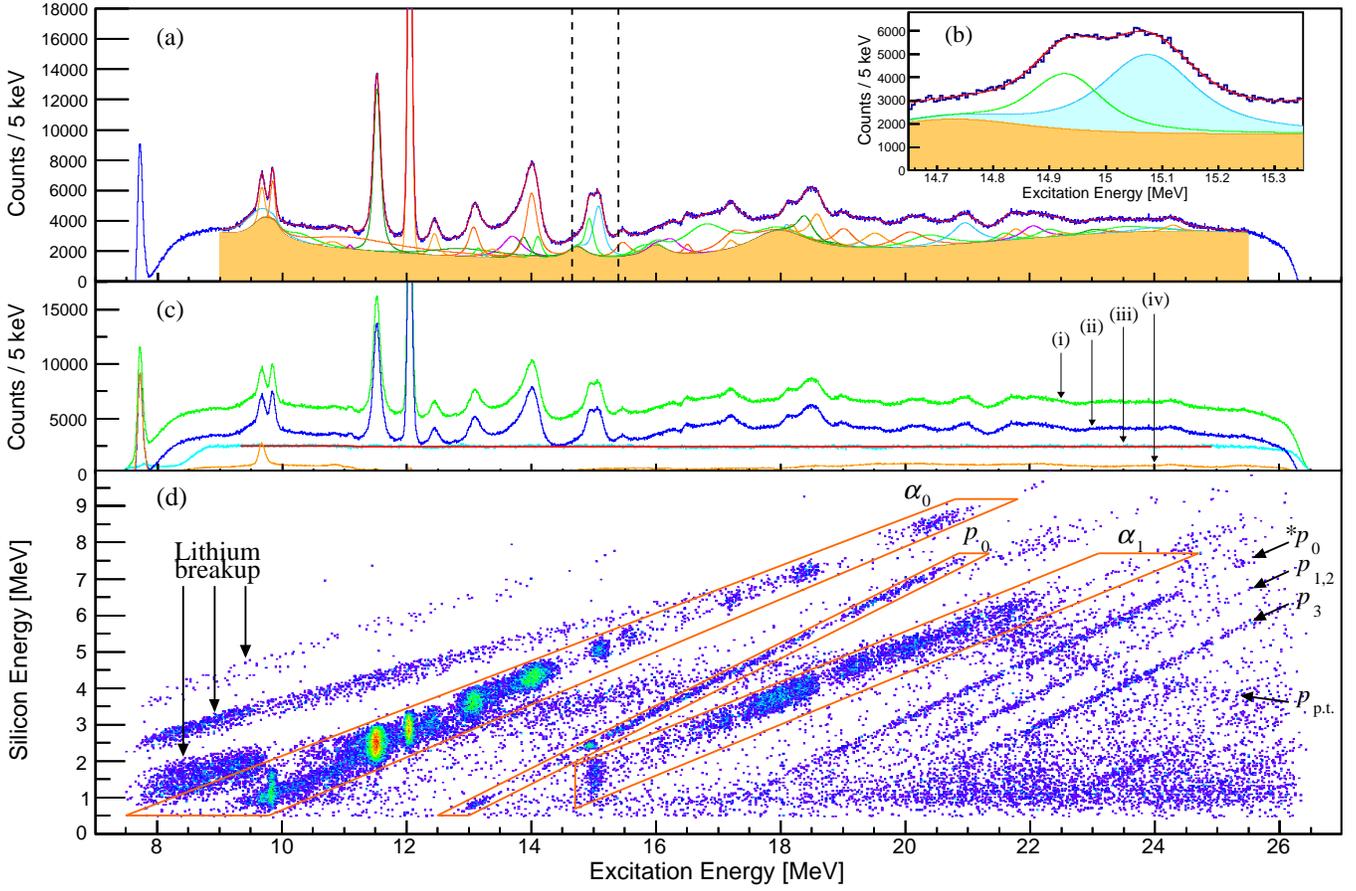}%
\caption{%
(Color online)
(a): The background subtracted inclusive excitation spectrum from the $\mathrm{Li_{2}CO_{3} }$ target, with fitted 
R-Matrix Voigt lineshapes which are superimposed upon the background of fitted lithium resonances (filled in orange). 
(b): The excitation energy region of interest highlighting the $\mathrm{0_{6}^{+} }$ resonance (blue line) and the 
neighbouring $J^{\pi}$= $2^{+}$ resonance (green line).
(c): The raw (i), background subtracted (ii) and instrumental background inclusive excitation spectra (iii) from the $\mathrm{Li_{2}CO_{3} }$ target. 
Spectrum (iv) is the background subtracted inclusive excitation spectrum from the $^{12}\mathrm{C}$ target, normalised 
to spectrum (b) through the 9.641(5) MeV resonance in $^{12}\mathrm{C}$. 
The linear fit of spectrum (c), used for background subtraction, is drawn in red.
(d): The coincident matrix of silicon energy versus the excitation energy of the recoil nucleus for decay particles detected within the angular range: $156^{\circ} < \theta_{lab} < 163^{\circ}$ (two strips within the array). 
The $\alpha_{0}$, $\alpha_{1}$ and $p_{\text{0--3}}$ decay lines from $^{16}\mathrm{O}$ are indicated. The proton punchthrough structure from the $p_{0}$ decay is labelled $p_{\textrm{p.t.}}$.
The lithium breakup and the *$p_{0}$ decay line from $^{12}\mathrm{C}$ are indicated.
A display color threshold of $>$1 is imposed.
}
\label{fig:/figs/spectra/Singles_SI}
\end{figure*}



Recent unsuccessful attempts to measure particle decay widths of the $0_{6}^{+}$ state in $\mathrm{^{16}O}$ \cite{Thomas2010, PhysRevC.83.064324} 
highlighted the need for an experiment that combines $\alpha$-particle decay measurements with a high energy resolution experimental setup and a reaction capable of preferentially populating $\mathrm{0^{+}}$ states.
In contrast to transfer reaction measurements,
inelastic $\alpha$-particle scattering at zero degrees has the advantage that it predominantly excites low-spin natural parity states. 
A measurement of the $\mathrm{^{16}O}(\alpha, \alpha^{\prime})$ reaction at zero degrees, coupled with coincident observations of the $\mathrm{^{16}O}$ decay products, was performed at the iThemba Laboratory for Accelerator-Based Sciences (iThemba LABS) in South Africa.
A 200-MeV dispersion-matched $\alpha$-particle beam was provided by the separated sector cyclotron. The $\alpha$-particles which were inelastically scattered off a $\mathrm{^{nat}Li_{2}CO_{3} }$ target were momentum-analyzed at zero degrees with the
K600 magnetic spectrometer \cite{Neveling-2011}. 
The energy resolution obtained was 85(1) keV FWHM, determined from the 12.049(9) MeV resonance in $\mathrm{^{16}O}$.
The error on the calculated excitation energy was $\delta E_{x} < 9$ keV.
The 510-$\mathrm{\mu g/cm^{2}}$-thick $\mathrm{^{nat}Li_{2}CO_{3}}$ target 
was prepared on a 50-$\mathrm{\mu g/cm^{2}}$-thick $\mathrm{^{12}C}$ backing.
The total Li content was approximately 50-$\mathrm{\mu g/cm^{2}}$ \cite{Papka2015}.
%
The solid-angle acceptance of the spectrometer (3.83 msr) was defined by a circular collimator with an opening angle $\pm2^{\circ}$. 
A comprehensive description of the experimental and analysis techniques is reported elsewhere \cite{KevinCWLi-Masters}.

The inclusive excitation energy spectra are displayed in Figure \ref{fig:/figs/spectra/Singles_SI}, panels (a), (b) and (c). 
In order to extract the excitation energy spectrum for the $\mathrm{^{16}O}(\alpha, \alpha^{\prime})$ reaction, 
the instrumental background must be taken into account and the contributions from the carbon and lithium present in the target must be identified.
The flat and featureless instrumental background, indicated as spectrum (iii) in Figure \ref{fig:/figs/spectra/Singles_SI} (c), 
is typical for measurements at zero degrees.
It results from small-angle elastic scattering off the target foil that is followed by re-scattering off 
any exposed part inside the spectrometer \cite{Neveling-2011}.
In order to subtract this background contribution, the spectrometer was operated in focus mode where 
the quadrupole at the entrance to the spectrometer was used to vertically focus reaction products to a narrow horizontal band on the focal plane. 
The background spectrum was generated by using the sections of the focal plane above and below the focused vertical band.
A linear fit was employed to approximate the background and was subtracted from the raw spectrum (i) to produce spectrum (ii). 
The background subtracted focal plane spectrum from the $\mathrm{^{12}C}$ target (iv) was normalised to spectrum (ii) 
through the 9.641(5) MeV resonance in $^{12}\mathrm{C}$. 
The smooth contribution observed from $\mathrm{^{12}C}$ in the excitation energy region of interest ($E_{x}\approx 15$ MeV), 
combined with the broad resonances of $\mathrm{^{7}Li}$ and $\mathrm{^{6}Li}$ 
indicated by the orange band in Figure \ref{fig:/figs/spectra/Singles_SI} (a) and (b), 
ensure that the distinctly observed resonances can be assigned to $\mathrm{^{16}O}$.
At $E_{x}\approx 15$ MeV, the decay modes from $^{16}\mathrm{O}$ are not affected by the lithium breakup indicated on Figure \ref{fig:/figs/spectra/Singles_SI} (d).

The decay products were observed with the Coincidence Array for K600 Experiments (CAKE) \cite{Adsley:2016blb}, consisting of 
four TIARA HYBALL MMM-400 double sided silicon strip detectors (DSSSDs). Each of the 400 $\mu$m thick 
wedge-shaped DSSSDs consists of 16 rings and 8 sectors, and were positioned at backward angles with the 
rings covering the polar-angle range of $114^{\circ} < \theta_{lab} < 166^{\circ}$, resulting in a solid-angle coverage 
of 21\% of $4\pi$. 
For each focal-plane event, all signals from CAKE within a time window of 6 $\mu$s were digitized, yielding both K600 inclusive 
as well as K600 + CAKE coincidence events. A beam pulse selector at the entrance of the cyclotron (which accepted 1 in 5 pulses) was 
employed to ensure a sufficient time window (273 ns) for coincidence measurements.

\begin{figure}[h]
\centering
\includegraphics[width=0.5\textwidth]{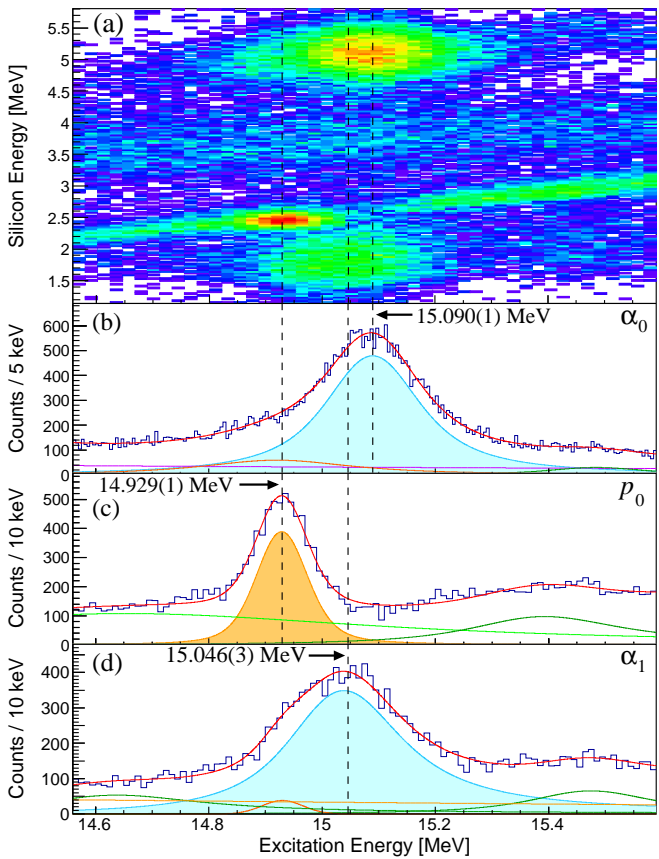}
\caption{(Color online) (a) The coincident matrix of silicon energy versus the excitation energy of the recoil nucleus, highlighting the decay modes observed at the excitation energy region of interest at $E_{x}\approx 15.1$ MeV.
The excitation energy projections of the $\alpha_{0}$, $p_{0}$ and $\alpha_{1}$ decay lines are displayed in (b), (c) and (d) respectively. The resonance energies, extracted with single channel R-matrix fits, are indicated.}
\label{fig:/figs/spectra/alpha0_alpha1_p0_FIT}
\end{figure}

The detection of coincident charged decay particles with the CAKE enables the characterization of resonances through the 
measurement of 
branching ratios and angular correlations of various decay modes.
The associated decay lines of the $\mathrm{^{16}O}$ nucleus are displayed in Figure \ref{fig:/figs/spectra/Singles_SI} (d): 
$\alpha$-decay and proton decay to the ground state of the residual nucleus are designated $\alpha_{0}$ and $p_{0}$ respectively, 
whilst $\alpha$-decay to the first excited state is designated $\alpha_{1}$.
By gating upon a particular decay line and projecting onto excitation energy, the resonance lineshape corresponding to a 
particular decay mode can be observed in isolation. 
The excitation energy spectra around $E_{x}\approx 15$ MeV corresponding to the $\alpha_{0}$, $p_{0}$ and $\alpha_{1}$ decay modes 
are displayed in Figure \ref{fig:/figs/spectra/alpha0_alpha1_p0_FIT} on panels (b), (c) and (d) respectively.

Resonances in the energy range of interest exhibit an R-matrix Lorentzian lineshape:

\begin{equation}
   \centering
	N(E) \propto \frac{\Gamma(E)}{\left[E - E_{R}\right]^{2} + \left[\Gamma(E)/2\right]^{2}},
   \label{eq:Lineshape}
\end{equation}

\noindent where $E_{R}$ is the resonance energy (location parameter) and the total width, $\Gamma(E)$, is a sum of the energy-dependent partial widths. 
For the \textit{$i^{th}$} decay channel, the partial width is given by

\begin{equation}
   \centering
	\Gamma_{i}(E) = 2\gamma_{i}^{2}P_{l}(E), 
   \label{eq:R-Matrix Resonance Width}
\end{equation}

\noindent where $\gamma_{i}$ is the reduced width and $P_{l}(E)$ is the associated penetrability, corresponding to the orbital angular momentum of decay $l$ and the chosen external radius given by $R = 1.2 \times (A_{1}^{1/3} + A_{2}^{1/3})$. 
Given the inherent resolution of the focal plane detector system, the experimentally observed lineshape of a resonance takes the form of a convolution between a Gaussian and the aforementioned R-matrix Lorentzian lineshape, approximated by a Voigt lineshape \cite{Wells199929} with an R-matrix energy-dependent width, $\Gamma(E)$.

A single-channel R-Matrix fit was implemented across the entire range of the focal plane considering possible 
resonances from all four target nuclei: $\mathrm{^{16}O}$, $\mathrm{^{12}C}$, $\mathrm{^{7}Li}$ and $\mathrm{^{6}Li}$. 
The Voigt lineshapes within the fit were assigned the experimental energy resolution of $85(1)$ keV.
%
%
For all fitted resonances, the decay parameters of each lineshape were chosen to correspond to the decay channel with the lowest orbital angular momentum.
For each resonance with unknown branching ratios, the lineshape was prescribed the decay channel parameters corresponding to the most strongly observed decay mode of the resonance (from this work).
The fitted resonance energy, $E_{R}$, of each known resonance was constrained to within $3\sigma$ about its literature value 
(excluding the resonances at $E_{x}\approx 15$ MeV).
An upper limit on the width, known as the Wigner limit \cite{PhysRev.98.145}, was imposed on each decay channel.
The extracted total width of each resonance, $\Gamma(E)$, is evaluated at the associated resonance energy $E_{R}$ 
(see Equation \ref{eq:R-Matrix Resonance Width}).
In the inclusive spectrum displayed on Figure \ref{fig:/figs/spectra/Singles_SI} (a) and (b), a prominent resonance was observed 
at $E_{x} = 15.076(7)$ MeV with an associated width of 162(4) keV.
This is in good agreement with previous measurements of the $0_{6}^{+}$ resonance, as displayed in Table \ref{fig:/table/PreviousResults}. 
In contrast, the observed width of 101(3) keV for the neighbouring $J^{\pi}$= $2^{+}$ resonance at $E_{R} = 14.926(2)$ MeV
does not agree well with the corresponding literature value of 54(5) keV. 
By fitting the focal plane spectra gated on the $\alpha_{0}$, $p_{0}$ and $\alpha_{1}$ decay modes, 
the resonance energies and widths from the resonances around $E_{x} \approx15$ MeV were extracted, 
as displayed in Table \ref{fig:/table/ExtractedResults}.

\begin{figure}[h!]
\centering
\includegraphics[width=0.5\textwidth]{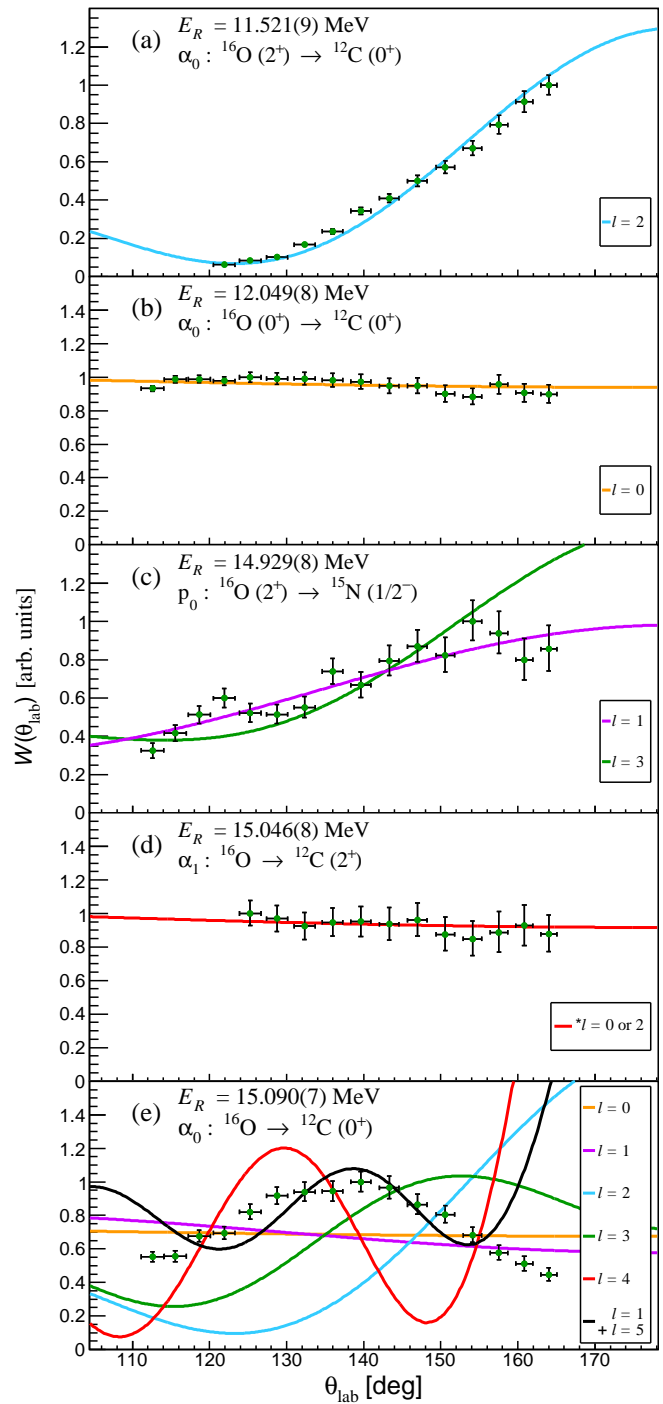}
\caption{(Color online) Angular correlations of charged particle decays from $^{16}\mathrm{O}$ in the laboratory frame relative to the beam axis: %
(a) $\alpha_{0}$ decay from the 11.521(9) MeV $J^{\pi}$= $2^{+}$ resonance, %
(b) $\alpha_{0}$ decay from the 12.049(8) MeV $J^{\pi}$= $0^{+}$ resonance, %
(c) $p_{0}$ decay from the 14.929(8) MeV $J^{\pi}$= $2^{+}$ resonance, %
(d) $\alpha_{1}$ decay observed at 15.046(8) MeV, %
(e) $\alpha_{0}$ decay observed at 15.090(7) MeV.
*The $\alpha_{0}$ decays from $J^{\pi}$= $0^{+}$ and $2^{+}$ resonances, corresponding to $l = 0\hbar$ and $2\hbar$ respectively, exhibit the same angular correlations. Data points affected by the electronic thresholds of the CAKE array are omitted.
} %
\label{fig:/figs/spectra/AngDistFig_R3}
\end{figure}

%
By gating on events detected in particular rings in the CAKE array, angular correlations of decay can be 
extracted in the laboratory reference frame, as shown in Figure \ref{fig:/figs/spectra/AngDistFig_R3}.
Self-consistency of the R-matrix fits for the angular correlations was achieved by fixing the 
resonance energies and widths to the values extracted from the total fit of the relevant decay mode.
In order to calculate the theoretical angular correlations of decay in the laboratory frame, 
the differential cross sections for the population of natural parity states through the 
$\mathrm{^{16}O}(\alpha, \alpha^{\prime})$ reaction were calculated in the distorted-wave Born approximation with the code CHUCK3 \cite{CHUCK3_Kunz}.
Both the $m$-state population ratios and the angular correlations of subsequent particle decay from 
the recoil nucleus were then calculated with ANGCOR \cite{AngCor_Harakeh_Put} in the inertial reference frame of the recoil nucleus.
Considering the $\mathrm{^{16}O}(\alpha, \alpha^{\prime})$ reaction with an incident energy of $\textrm{E}_{lab}$ = 200 MeV 
and a recoil excitation energy of $E_{x} = 15.0$ MeV, the angular acceptance of the ejectile $\alpha$-particle corresponds to recoil nuclei ($^{16}\mathrm{O}$) with minimum and maximum kinetic energies of 80 keV ($\theta_{lab}=0^{\circ}$) and 140 keV ($\theta_{lab}=40^{\circ}$) respectively.
To account for the velocity of the recoil nuclei, the calculated correlations are then relativistically 
transformed to the laboratory frame by taking into account the angular acceptance for the ejectile nuclei by the spectrometer.
The solid-angle correction factors for the CAKE array are obtained with a GEANT4 simulation 
which accounts for a potential
$\pm2$ mm positioning error that has been propagated through to the total error of the data points.
Calculated angular correlations are shown in Figure \ref{fig:/figs/spectra/AngDistFig_R3}.

\begin{table}[b]
\begin{ruledtabular}
\begin{tabular}{ l  l  l  l l}
    \multirow{2}{*}{$J^{\pi}$}			&	Decay			&	$E_{R}$			&	$\Gamma_{total}$	&	Branching				\\
								&	Mode			&	{[}MeV{]} 			&	{[}MeV{]}			&	Ratio	 [\%]			\\
   \hline
\rule{0pt}{9.5pt}	\multirow{2}{*}{$2^{+}$}	&	Inclusive			&	11.520(9)			&	80(1)				& -		\\
\rule{0pt}{9.5pt}						&	$\alpha_{0}$		&	11.521(9)			&	82(1)				& 109(3)			\\
\hline
\rule{0pt}{9.5pt}	\multirow{2}{*}{$0^{+}$}	&	Inclusive			&	12.049(9)			&	5(1)				& -				\\
\rule{0pt}{9.5pt}						&	$\alpha_{0}$		&	12.049(8)			&	12(1)				& 96(3)$^\dagger$			\\
\hline
\rule{0pt}{9.5pt}	\multirow{2}{*}{$2^{+}$}	&	Inclusive			&	14.930(8)			&	101(3)			& -				\\
\rule{0pt}{9.5pt}						&	$p_{0}$			&	14.929(8)			&	40(1)				& 21(1)			\\
\hline
\rule{0pt}{9.5pt}	\multirow{3}{*}{$0^{+}$}	&	Inclusive			&	15.076(7)			&	162(4)			& -				\\
\rule{0pt}{9.5pt}						&	$\alpha_{0}$		&	15.090(7)			&	162(4)			& 72(2)$^\ddagger$			\\
\rule{0pt}{9.5pt}						&	$\alpha_{1}$		&	15.046(8)			&	216(10)			& 67(3)$^\ddagger$				\\
\end{tabular}
\end{ruledtabular}
   \caption[Extracted single channel R-matrix fit parameters from the inclusive and coincidence spectra.]{Extracted single channel R-matrix fit parameters from the inclusive and coincidence spectra. $^{\dagger}$The $\alpha_{1}$ decay channel was not observable within this work due to electronic thresholds of the CAKE array. $^{\ddagger}$The branching ratios are calculated under the assumption that the associated resonance is of $0^{+}$ spin and parity.}
   \label{fig:/table/ExtractedResults}
\end{table}



The data suggest the presence of a previously unidentified resonance at $E_{x}\approx15$ MeV.  
%
Figure \ref{fig:/figs/spectra/AngDistFig_R3} (e) shows that the angular distribution of $\alpha_{0}$ decay observed 
at $E_{R} = 15.090(7)$ MeV is distinctly anisotropic. This can only result from the presence of a previously 
unidentified resonance with non-zero spin, which may be obscuring the isotropic decay of the $J^{\pi} = 0_{6}^{+}$ resonance.
The calculated angular correlations of $\alpha_{0}$ decay from $J^{\pi} = 0^{+}, 1^{-}, 2^{+}, 3^{-}, 4^{+}, 5^{-}$ resonances 
at this resonance energy do not fit well to the data.
The two best fitting $\alpha_{0}$ angular correlations correspond to a $J^{\pi} = 0^{+}$ and a $J^{\pi} = 1^{-}$ resonance, 
yielding $\goodchi^{2}_{\textrm{red}}$ = 13.54 and $\goodchi^{2}_{\textrm{red}}$ = 16.65 respectively.
Given the possibility of two distinct but unresolved resonances, all possible pairs of the calculated correlations 
were incoherently summed and fitted to the data (with the relative contributions being free parameters). 
These calculations do not yield satisfactory reproduction of the data: the best fit corresponds to the incoherently summed $\alpha_{0}$-decay distributions from a $J^{\pi} = 1^{-}$ and a $J^{\pi} = 5^{-}$ resonance, yielding $\goodchi^{2}_{\textrm{red}}$ = 7.67. 
It is possible that the angular correlations of these inherently overlapping resonances may interfere.
The anisotropy remains a clear identifier of a resonance with non-zero spin.
To ensure that the anisotropy is not a consequence of the analysis, the angular correlations of decay from the 
most prominently observed resonances are also analyzed.
The $\alpha_{0}$ angular distribution of the $J^{\pi} = 0^{+}$ resonance at \mbox{$E_{x} = 12.049(9)$ MeV}, shown in Figure \ref{fig:/figs/spectra/AngDistFig_R3} (b), exhibits isotropy and the corresponding calculation fits the data with a reduced chi-squared of \mbox{$\goodchi^{2}_{\textrm{red}}$ = 1.01}, indicating that the experimental setup is well understood.
Similarly for the $\alpha_{0}$ angular distribution of the $J^{\pi} = 2^{+}$ resonance at \mbox{$E_{x} = 11.520(9)$ MeV} 
displayed in Figure \ref{fig:/figs/spectra/AngDistFig_R3} (a), the theoretical fit is reasonable and
yields $\goodchi^{2}_{\textrm{red}}$ = 1.42.

The angular distribution of the $\alpha_{1}$ decay mode observed at \mbox{15.046(8) MeV} is observed to be isotropic, 
as displayed on Figure \ref{fig:/figs/spectra/AngDistFig_R3} (d).
Whilst only a $J^{\pi} = 0^{+} \textrm{or } 2^{+}$ resonance can exhibit purely isotropic $\alpha_{1}$ decay, inherently anisotropic 
decays from resonances of other spin-parities may experimentally appear isotropic given their multiple possible $l$-values of decay. 
It is therefore assumed that this $\alpha_{1}$ decay mode originates from the $J^{\pi} = 0_{6}^{+}$ resonance.
The fit to the calculated $\alpha_{1}$ $J^{\pi} = 0^{+}$ angular distribution yields $\goodchi^{2}_{\textrm{red}}$ = 0.15, which could indicate either an overestimation of the data errors, or that the dominant error is a systematic scaling factor. This, however, does not affect the conclusions of the paper.
%
The angular distribution of the $p_{0}$ decay mode observed at \mbox{14.929(8) MeV} is displayed on Figure \ref{fig:/figs/spectra/AngDistFig_R3} (c).
The proton-decay from a $2^{+}$ resonance to the $1/2^{-}$ ground state of $\mathrm{^{15}N}$ corresponds to orbital angular momenta of decay of either $l = 1\hbar$ or $3\hbar$, corresponding to calculated correlations which fit with $\goodchi^{2}_{\textrm{red}}$ = 1.25 and 4.57 respectively.
%

Additional evidence towards a previously unidentified resonance is given by the extracted resonance energies 
from the fitted lineshapes for the $\alpha_{0}$ and $\alpha_{1}$ decay modes.
The resonance energies corresponding to various decay modes from a resonance can provide insight into the 
spin and parity, particularly when only a single $l$-value of decay is possible.
For a $J^{\pi}$= $0^{+}$ resonance in $^{16}\mathrm{O}$, an $\alpha$-particle emitted either through 
the $\alpha_{0}$ or $\alpha_{1}$ decay carries exactly $l=0\hbar$ or $2\hbar$ units of angular momentum respectively. 
%
If the $\alpha_{0}$ and $\alpha_{1}$ decay modes observed at $E_{x} \approx15$ MeV are from the same $J^{\pi}$= $0^{+}$ resonance in $\mathrm{^{16}O}$, 
both the greater centre-of-mass energy and the lack of a centrifugal potential barrier for $\alpha_{0}$ decay
suggest that extracted resonance energy of the $\alpha_{0}$ decay lineshape should be lower than that of the $\alpha_{1}$ decay.
%
%
From this work, the $\alpha_{0}$ decay mode observed at $E_{x} = 15.090(7)$ MeV is 44(3) keV higher in excitation energy than that of the $\alpha_{1}$ decay mode.
It is therefore incompatible for the $\alpha_{0}$ and $\alpha_{1}$ decay modes to both originate from a single $J^{\pi}$= $0^{+}$ resonance.
In principle, this shift in resonance energies could be explained by the existence of either a single $J^{\pi}$= $2^{+}$ or $J^{\pi}$= $3^{-}$ resonance: the minimal orbital angular momenta for $\alpha_{0}$ and $\alpha_{1}$ decay are $l=2\hbar$ or $0\hbar$ respectively for a $J^{\pi}$= $2^{+}$ resonance and $l=3\hbar$ or $1\hbar$ respectively for a $J^{\pi}$= $3^{-}$ resonance.
Assuming the \textit{m}-state population ratios calculated with a direct single-step reaction mechanism are correct, the calculated angular correlations of $\alpha_{0}$ decay from both a $J^{\pi}$= $2^{+}$ and a $J^{\pi}$= $3^{-}$ resonance do not agree well with the data displayed in Figure \ref{fig:/figs/spectra/AngDistFig_R3} (e).

Finally, we note that the presence of a previously unidentified resonance at $E_{x}\approx15$ MeV
could explain why the extracted total width of the unresolved $J^{\pi}$= $2^{+}$ resonance at $E_{x} = 14.930(8)$ MeV, extracted from the inclusive data to be 101(3) keV, is inconsistent with the $p_{0}$-extracted width 
and literature value of 40(1) keV and 54(5) keV respectively.
The observation of a smooth and featureless instrumental background spectrum (iii) in Figure \ref{fig:/figs/spectra/Singles_SI} (c) 
ensures that this disparity of widths is not caused by experimental artefacts.
Similarly, the inclusive excitation energy spectrum from the $\mathrm{^{12}C}$ target, 
displayed as spectrum (iv) in Figure \ref{fig:/figs/spectra/Singles_SI} (c), 
shows that the $\mathrm{^{12}C}$ contribution at $E_{x}\approx15$ MeV is negligible.
Given the $\alpha$-separation energies for $\mathrm{^{6}Li}$ and $\mathrm{^{7}Li}$ of $E_{sep} = 1.47 \textrm{ and } 2.47$ MeV respectively, 
the contributions of the lithium resonances to the focal plane spectra collectively form 
a slowly-varying continuum, shown as the orange-filled lineshape in Figure \ref{fig:/figs/spectra/Singles_SI} (a) and (b). 
Furthermore, the presence of a contaminant nucleus which decays through charged-particle emission 
would be kinematically identified within the coincident matrix of silicon energy versus excitation energy, 
displayed in Figure \ref{fig:/figs/spectra/Singles_SI} (d).

Itoh {\it et al.} studied the $\mathrm{^{16}O}(\alpha, \alpha^{\prime})$ reaction at $\theta_{lab} = 0^\circ$ and $\theta_{lab} = 4^\circ$, 
with an incident energy of \mbox{$\textrm{E}_{lab}$ = 386 MeV} \cite{1742-6596-569-1-012009}. 
A multipole decomposition was performed on the differential cross section of the resonance within the excitation 
energy interval: \mbox{$15.00 \textrm{ MeV} < E_{x} < 15.25 \textrm{ MeV}$}. Whilst the decomposition indicated the presence of a $0^{+}$ resonance, the differential cross section is qualitatively different to that of the $0^{+}$ resonance observed 
at \mbox{$12.00 \textrm{ MeV} < E_{x} < 12.25$} MeV. This is reflected by the larger fitted contribution of $L\ge1$ angular 
momentum transfer reactions. Their work is therefore consistent with the existence of a previously 
unresolved resonance at $E_{x}\approx15$ MeV which does not exhibit a $0^{+}$ nature.

By studying the $\mathrm{^{16}O}(\alpha, \alpha^{\prime})$ reaction at $\theta_{lab} = 0^\circ$ with an 
incident energy of \mbox{$\textrm{E}_{lab}$ = 200 MeV}, low spin states in $\mathrm{^{16}O}$ were strongly excited. 
The angular correlations observed with the CAKE suggest the existence of a previously unresolved resonance 
at $E_{x}\approx15$ MeV with non-zero spin.
This is supported by the shift in resonance energies between the $\alpha_{0}$ and $\alpha_{1}$ decay modes 
(see Figure \ref{fig:/figs/spectra/alpha0_alpha1_p0_FIT}). 
The existence of a previously unresolved resonance may explain the disparity between the 
theoretical and experimentally observed widths of 34 keV and 160(30) keV respectively.


This work was supported by the South Africa National Research Foundation, and in particular, through NEP grant 86052. 
RN acknowledges financial support from the NRF through grant number 85509.


\bibliography{apssamp}

\end{document}